\pdfoutput=1
 \documentclass[aps,prl,reprint,twocolumn,showpacs,superscriptaddress,nofootinbib]{revtex4-1}  

\usepackage{tabularx}
\makeatletter
\def\hlinewd#1{%
\noalign{\ifnum0=`}\fi\hrule \@height #1 %
\futurelet\reserved@a\@xhline}
\makeatother
\usepackage{scrextend}
\usepackage{amsmath}
\usepackage{amssymb}
\usepackage{epsfig}
\usepackage{graphicx}
\usepackage{hyperref}
\usepackage{dcolumn}   
\usepackage{slashed}
\usepackage{color}
\usepackage{rotating}
\usepackage[margin=0.9in,a4paper]{geometry}
\usepackage[table,xcdraw,dvipsnames]{xcolor}
\usepackage[utf8]{inputenc}
\usepackage{colortbl}
\usepackage[normalem]{ulem}

\definecolor{nicered}{rgb}{0.7,0.1,0.1}
\definecolor{nicegreen}{rgb}{0.1,0.5,0.1}
\definecolor{red}{rgb}{1.0, 0, 0}

\newcommand{\X}{{\cal X}}

\renewcommand{\P}{{\cal P}}

\newcommand{\SU}{{\rm SU}}

\newcommand{\U}{{\rm U}}
\newcommand\e{\mathrm e}

\renewcommand{\[}{\left[}

\renewcommand{\(}{\left(}
\renewcommand{\)}{\right)}

\newcommand{\bdm}{\begin{displaymath}}
\newcommand{\edm}{\end{displaymath}}
\newcommand{\bea}{\begin{eqnarray}}
\newcommand{\eea}{\end{eqnarray}}

\newcommand{\nn}{\nonumber}
\renewcommand{\S}{\mathcal{S}}
\renewcommand{\X}{\mathcal{X}}

\renewcommand\P{\mathcal P}
\renewcommand{\a}{\alpha}
\renewcommand{\b}{\beta}
\renewcommand{\t}{\theta}

\definecolor{nicered}{rgb}{0.7,0.1,0.1}
\definecolor{nicegreen}{rgb}{0.1,0.5,0.1}
\definecolor{red}{rgb}{1.0, 0, 0}
\definecolor{niceblue}{rgb}{0,0,0.8}
\definecolor{red}{rgb}{1.0, 0, 0}
\hypersetup{colorlinks,citecolor= nicegreen,linkcolor= nicered,urlcolor=nicered}

\def\eq#1{{Eq.~(\ref{#1})}}
\def\eqs#1#2{{Eqs.~(\ref{#1})--(\ref{#2})}}

\def\fig#1{{Fig.~\ref{#1}}}

\def\vev#1{\left\langle #1\right\rangle}

\def\diag{\mbox{diag}\,}

\def\gsim{\raise0.3ex\hbox{$\;>$\kern-0.75em\raise-1.1ex\hbox{$\sim\;$}}}
\def\lsim{\raise0.3ex\hbox{$\;<$\kern-0.75em\raise-1.1ex\hbox{$\sim\;$}}}

\def\mb[#1]{\mathbf{#1}}

\renewcommand{\bar}{\overline}

\definecolor{LightCyan}{rgb}{0.88,1,1}
\definecolor{piggypink}{rgb}{0.99, 0.87, 0.9}
\definecolor{applegreen}{rgb}{0.55, 0.71, 0.0}
\definecolor{darkpastelgreen}{rgb}{0.01, 0.75, 0.24}
\definecolor{green-yellow}{rgb}{0.68, 1.0, 0.18}

\newcommand{\beq}{\begin{equation}}
\newcommand{\eeq}{\end{equation}}
\newcommand{\beqa}{\begin{eqnarray}}
\newcommand{\eeqa}{\end{eqnarray}}
\newcommand{\eps}{\varepsilon}

\newcommand{\Sec}[1]{ \medskip \noindent {\sl \bfseries #1}}

\begin{document}



\title{
Axion-mediated forces, CP violation and left-right interactions
} 

\author{Stefano Bertolini}
\email{stefano.bertolini@sissa.it}
\affiliation{\normalsize \it 
INFN, Sezione di Trieste, SISSA, Via Bonomea 265, 34136 Trieste, Italy}
\author{Luca Di Luzio}
\email{luca.diluzio@desy.de}
\affiliation{\normalsize \it 
Deutsches Elektronen-Synchrotron 
DESY, Notkestra\ss e 85, 
D-22607 Hamburg, Germany}
\author{Fabrizio Nesti}
\email{fabrizio.nesti@aquila.infn.it}
\affiliation{\normalsize \it 
  Dipartimento di Scienze Fisiche e Chimiche, Universit\`a dell'Aquila, via Vetoio, I-67100, L'Aquila, Italy}
\affiliation{\normalsize \it INFN, Laboratori Nazionali del Gran Sasso, I-67100 Assergi (AQ), Italy}

\begin{abstract}
\noindent
We compute the CP-violating (CPV) scalar axion coupling to nucleons in the framework of baryon
chiral perturbation theory and we apply the results to the case of left-right symmetry.
The correlated constraints with other CPV observables show that the predicted axion nucleon coupling
is within the reach of present axion-mediated force experiments for $M_{W_R}$ up to $1000$ TeV.
\end{abstract}

\maketitle

\Sec{Introduction.}  The axion experimental program has received an impressive boost in the last
decade.  Novel detection strategies, bridging distant areas of physics, promise to open for
exploration the parameter space of the QCD axion in the not-so-far future, possibly addressing the
issue of strong CP violation in the Standard Model (SM) via the Peccei-Quinn (PQ) mechanism
\cite{Peccei:1977hh,Peccei:1977ur,Weinberg:1977ma,Wilczek:1977pj} and the Dark Matter (DM) puzzle
\cite{Preskill:1982cy,Abbott:1982af,Dine:1982ah} (for updated reviews, see
\cite{Sikivie:2020zpn,DiLuzio:2020wdo,Irastorza:2018dyq}).  Standard axion searches often rely on
highly model-dependent axion production mechanisms, as in the case of relic axions (haloscopes) or
to a less extent solar axions (helioscopes); while traditional optical setups in which the axion is
produced in the lab are still far from probing the standard QCD axion.  A different experimental
approach, as old as the axion itself \cite{Weinberg:1977ma}, consists in searching for
axion-mediated macroscopic forces \cite{Moody:1984ba}.  Given the typical axion Compton wavelength
$\lambda_a \sim 2 \, \text{cm} \, (10 \, \mu\text{eV} / m_a)$,
an even tiny
\emph{scalar} axion coupling to matter may coherently enhance the force between macroscopic
bodies.  The sensitivity of these experiments crucially depends on the (pseudo)scalar nature of the
axion field, a matter of ultraviolet (UV) physics.

Within QCD the Vafa-Witten theorem \cite{Vafa:1984xg} ensures that the axion vacuum expectation
value (VEV) relaxes on the $\bar{\theta}_{\rm eff} \equiv \vev{a} / f_a + \bar{\theta} = 0$ minimum,
where $\bar{\theta}$ denotes the QCD topological term.  However, extra CP violation in the UV
invalidate the hypotheses of this theorem, and in general one expects a minimum with
$\bar{\theta}_{\rm eff} \neq 0$.  While the CKM phase in the SM yields
$\bar{\theta}_{\rm eff} \simeq 10^{-18}$~\cite{Georgi:1986kr}, too tiny to be experimentally
accessible, CPV phases from new physics can saturate the neutron Electric Dipole Moment (nEDM) bound
$|\bar{\theta}_{\rm eff}| \lesssim 10^{-10}$.

Another remarkable consequences of a non-zero $\bar{\theta}_{\rm eff}$ is the generation of CPV
scalar axion couplings to nucleons, $\bar g_{aN}$, which is probed in axion-mediated force
experiments.  In particular, given the nEDM bound on $\bar{\theta}_{\rm eff}$ the
scalar-pseudoscalar combination (also known as monopole-dipole interaction) offers the best chance
for detecting the QCD axion.  Additionally, the presence of a spin-dependent interaction allows to
use Nuclear Magnetic Resonance (NMR) to enhance
the signal.  This is the strategy pursued by ARIADNE \cite{Arvanitaki:2014dfa,Geraci:2017bmq} which
aims at probing the monopole-dipole force
via a sample of nucleon spins.  A similar approach is pursued by QUAX-$g_pg_s$
\cite{Crescini:2016lwj,Crescini:2017uxs}, using instead electron spins.  ARIADNE will probe
$|\bar{\theta}_{\rm eff}| \lesssim 10^{-10}$ for axion masses
$1 \lesssim m_a / \mu\text{eV} \lesssim 10^4$, a range highly motivated by DM.

In this Letter, we provide a coherent framework for computing the CPV scalar axion coupling to
nucleons in terms of new sources of CP violation beyond the SM.  This is done in the framework of
the baryon chiral Lagrangian that allows us to compute all contributions of meson
tadpoles and $\bar \theta_{\rm eff}$ at once, as well as isospin-breaking effects.  In comparison to
previous works~\cite{Moody:1984ba,Barbieri:1996vt,Pospelov:1997uv,Bigazzi:2019hav}, the
contributions of the pion tadpole induced by the QCD dipole operator was estimated in
\cite{Barbieri:1996vt} by naive dimensional analysis and in \cite{Pospelov:1997uv} using current
algebra techniques, while isospin breaking was considered in \cite{Bigazzi:2019hav} for
$\bar \theta_{\rm eff}$ without meson tadpoles.  Our result is general and can be systematically
applied to any bosonic representation of P and CP violating effective operators induced in
extensions of the SM.

We detail our approach in the case of effective operators from RH currents, and then apply the
results in the minimal Left-Right symmetric model (LRSM) endowed with a PQ symmetry and $\P$-parity
as LR symmetry.  This is an extremely predictive and motivated case for neutrino masses and
additional CP violation, with an active collider physics program~\cite{Senjanovic:2011zz}.  We build
on the approach detailed in Ref.~\cite{Bertolini:2019out}, which presented a study of the kaon CPV
observables $\varepsilon$, $\varepsilon'$ and the nEDM ($d_n$) in minimal LR scenarios. It was found
there that the embedding of a PQ symmetry relaxes the lower bound on the LR scale just at the upper
reach of the LHC.  In this work we show that the present search for the scalar axion coupling to
nucleons provides correlated and complementary constraints, with a sensitivity to the LR scale
stronger than other CPV observables. Remarkably, for a non-decoupled LR-scale we obtain a
lower-bound on the $\bar g_{aN}$ coupling, thus setting a target for present axion-mediated force
experiments.

\Sec{CPV axion couplings to matter.} 
Including both CP-conserving and CPV couplings,  
the axion effective Lagrangian with matter fields ($f=p,n,e$) reads
\beq
\label{eq:Laint1}
\mathcal{L}_{af} = 
C_{af} \frac{\partial_\mu a}{2 f_a} \bar f \gamma^\mu \gamma_5 f 
- \bar{g}_{af} \, a \bar f f
\,,
\eeq
where the first term can be rewritten in terms of a pseudoscalar density as
$- g_{af} \, a \bar f i \gamma_5 f$, with $g_{af} = C_{af} m_f / f_a$.  For protons and neutrons the
adimensional axion coupling coefficients are~\cite{diCortona:2015ldu}
\begin{align}
\label{eq:Cap}
\!\!\! 
C_{ap} & = -0.47(3) + 0.88(3) \, c_u - 0.39(2) \, c_d - K_a
\\[1ex]
\label{eq:Can}
\!\!\! 
C_{an} &= -0.02(3) + 0.88(3) \, c_d - 0.39(2) \, c_u - K_a
\, , 
\end{align}
where $K_{a} = 0.038(5) \, c_s +0.012(5) \, c_c + 0.009(2) \, c_b + 0.0035(4) \, c_t$, and where the
(model-dependent) axion couplings to quarks $c_q$ are defined via the Lagrangian term
$c_q \frac{\partial_\mu a}{2f_a} \bar q \gamma^\mu \gamma_5 q$.  The axion mass and decay constant
are related by
$m_a = 5.691(51) \( 10^{12} \ \text{GeV} / f_a \) \,
\text{$\mu$eV}$~\cite{Gorghetto:2018ocs,Borsanyi:2016ksw}.

The origin of the CPV scalar couplings to nucleons $\bar{g}_{aN}$ $(N=p,n)$ can be traced back to
sources of either PQ or CP violation. These generically lead to a remnant
$\bar{\theta}_{\rm eff} \neq 0$ which induces CPV couplings.
One finds for the isospin singlet component of the matrix element~\cite{Moody:1984ba}
\begin{align} 
\label{eq:fromthetaefftogaN}
\bar{g}_{aN} &= \frac{\bar{\theta}_{\rm eff}}{f_a} \frac{m_u m_d}{m_u + m_d}  \frac{\langle N | 
               \bar u u + \bar d d | N \rangle}{2}\, ,
\end{align}
where we included a $1/2$ factor missed in~\cite{Moody:1984ba}.
A shortcoming of \eq{eq:fromthetaefftogaN} is that CPV physics can induce not only
$\bar{\theta}_{\rm eff}$, but also shifts the chiral vacuum, inducing tadpoles for the $\pi^0$,
  $\eta_0$, $\eta_8$ meson fields. These in turn yield extra contributions to $\bar{g}_{aN}$, as to
other CPV observables such as $d_n$.  A derivation of $g_{an,p}$
taking all these effects consistently into account is here obtained in the context of the baryon chiral Lagrangian with axion field, 
as described below. We find
\bea
\label{gan-pVEV}
\bar{g}_{an,\,p} &\simeq 
& \frac{4B_0\, m_u m_d}{f_a (m_u+m_d)}  \bigg[\pm (b_D+b_F)\frac{\vev{\pi^0}}{F_\pi}
 \\
 &&{}+ \frac{b_D-3b_F}{\sqrt{3}}\frac{\vev{\eta_8}}{F_\pi}  
 -\sqrt{\frac{2}{3}}(3b_0+2b_D) \frac{\vev{\eta_0}}{F_\pi}
\nn \\
&&   -   \left(b_0 + (b_D+b_F)\frac{m_{u,d}}{m_d+m_u}\right)\bar{\theta}_{\rm eff}\bigg] \, , \nn 
\eea
where for clarity we neglected $m_{u,d}/m_s$ terms. Here, $B_0=m_\pi^2/(m_d+m_u)$ while the hadronic
Lagrangian parameters $b_{D,F}$ are determined from the baryon octet mass splittings,
$b_D\simeq 0.07\,\rm GeV^{-1}$, $b_F\simeq -0.21\,\rm GeV^{-1}$ at LO~\cite{Pich:1991fq}.  The value
of $b_0$ is determined from the pion-nucleon sigma-term as $b_0\simeq -\sigma_{\pi N}/4m_\pi^2$.
From the precise determination in~\cite{Hoferichter:2015dsa,Hoferichter:2016ocj} one obtains
$b_0\simeq -0.76\pm 0.04\, \rm GeV^{-1}$ at 90\% C.L.
Given $\sigma_{\pi N}\equiv \langle N | \bar u u + \bar d d | N \rangle\, (m_u+m_d)/2$, the isospin symmetric  $b_0 \bar{\theta}_{\rm eff}$ term reproduces exactly \eq{eq:fromthetaefftogaN}.

\eq{gan-pVEV} represents our general result, including isospin-breaking effects, where
$\bar{\theta}_{\rm eff}$ and the meson VEVs are meant to be computed from a given source of
CPV. In general $\bar{g}_{aN}$ and $d_n$ are not proportional, as it would follow
  from~\eq{eq:fromthetaefftogaN}.  Exact cancellations among the VEVs can happen
for $d_n$~\cite{Cirigliano:2016yhc,Bertolini:2019out}.

\Sec{Axion coupling and RH currents.}
As a paradigmatic application, we explicitly compute the above CPV axion-matter coupling in the case
of RH currents, which arise in a wide class of models beyond the SM. Heavy RH currents lead
generally to four quark operators that violate P and CP as
${\cal O}_{1}^{qq'} = (\bar{q}q) \ (\bar{q'}i\gamma_5q')$,
$q=u,d,s$~\cite{Bertolini:2019out,An:2009zh,deVries:2012ab,Cirigliano:2016yhc,Haba:2018byj}.  Such
operators induce meson tadpoles and allow for a non-vanishing correlator with the topological
$G\tilde G$ term, thus shifting both chiral and axion vacua~\cite{Pospelov:1997uv}.  At the leading
order in momentum expansion the operators ${\cal O}_{1}^{qq'} $ are represented in the low-energy
meson Lagrangian by combinations of $[U^\dagger]_{qq}[U]_{q'q'}$ terms, where the usual
$3\times 3$ matrix $U$ represents nonlinearly the meson nonet under $U(3)_L\times U(3)_R$ rotations.
By a proper $U(3)_A$ field rotation, the axion field is also included in the meson and baryon chiral
Lagrangians.  Complete notation and details are found in Appendix D of~\cite{Bertolini:2019out}.
Rotating away the axion and meson tadpoles, the new CPV
axion-nucleon scalar couplings of \eq{gan-pVEV} are induced from the baryon Lagrangian.

In LR effective setups the operator ${\cal O}_{1}^{ud}$ generates typically the leading
contribution to $d_n$.  We show in this work that it also generates the dominant contribution to
$\bar g_{ap,n}$.  We denote its low scale Wilson coefficient as $C_1^{ud}$, and similarly for other
flavors.
When $O_1^{ud}$ is considered we find~\cite{Haba:2018byj,An:2009zh,Bertolini:2019out},
\bea
\frac{\langle\pi^0\rangle}{F_\pi}&\simeq&
\frac{G_F}{\sqrt{2}}\,{\cal C}_1^{[ud]}\,\frac{c_3}{B_0F_\pi^2}
\frac{m_u+m_d+4m_s}{m_um_d+m_dm_s+m_sm_u} \nn
\\[.5ex]
\frac{\langle\eta_8\rangle}{F_\pi}&\simeq& \frac{G_F}{\sqrt{2}}\,{\cal
  C}_1^{[ud]}\,\frac{\sqrt{3}c_3}{B_0F_\pi^2} \frac{m_d-m_u}{m_um_d+m_dm_s+m_sm_u}
\nn \\[.8ex]
\bar{\theta}_{\rm eff}&\simeq& \frac{G_F}{\sqrt{2}}\,{\cal C}_1^{[ud]}\,\frac{2c_3}{B_0F_\pi^2}
\frac{m_d-m_u}{m_um_d}\,,
\label{PQvev12}
\eea
where $ {\cal C}_1^{[ud]}\equiv{\cal C}_1^{ud}-{\cal C}_1^{du}$ and $\vev{\eta_0}=0$.  The
axion VEV  no longer cancels the original $\bar{\theta}$ term, leaving a calculable
$\bar{\theta}_{\rm eff}$.  As expected, the pion VEV is isospin odd ($u\leftrightarrow d$), while
the other VEVs are even.  The low-energy constant $c_{3}$ is estimated in the large $N$ limit as
$c_3\sim F_\pi^4 B_0^2/4$.  Another estimate, based on $\SU(3)$ chiral symmetry is given in
\cite{Cirigliano:2016yhc}.  Analogously, for ${\cal O}_1^{us}$ we find
\bea
\frac{\langle\pi^0\rangle}{F_\pi}&\simeq& \frac{G_F}{\sqrt{2}}\,{\cal
  C}_1^{[us]}\,\frac{c_3}{B_0F_\pi^2} \frac{2 m_d + 2 m_s - m_u}{m_um_d+m_dm_s+m_sm_u}
\nn \\[.7ex]
\frac{\langle\eta_8\rangle}{F_\pi}&\simeq& \frac{G_F}{\sqrt{2}}\,{\cal
  C}_1^{[us]}\,\frac{\sqrt{3}c_3}{B_0F_\pi^2} \frac{2 m_d + m_u}{m_um_d+m_dm_s+m_sm_u}
\nn \\[.7ex]
\bar{\theta}_{\rm eff}&\simeq& \frac{G_F}{\sqrt{2}}\,{\cal C}_1^{[us]}\,\frac{2c_3}{B_0F_\pi^2}
\frac{m_s-m_u}{m_um_s}\,.
\label{PQvev13}
\eea
One notices in both \eqs{PQvev12}{PQvev13} the $m_s/m_d$ enhancement of $\vev{\pi^0}$ over the other meson
VEV.

As observed in \cite{Cirigliano:2016yhc} and \cite{Bertolini:2019out}, the CPV coupling
$\bar{g}_{np\pi}$ computed using the VEVs (\ref{PQvev12}) vanishes identically.  On the other hand,
when ${\cal O}_1^{us}$ is considered, $\bar{g}_{n\Sigma^-K^+}$ cancels in turn.  In either case the
meson VEVs cancel exactly against $\bar{\theta}_{\rm eff}$, a result which is made transparent
in the basis of Ref.~\cite{Pich:1991fq}.

Such a cancellation is not present for the CPV axion-nucleon couplings $\bar g_{an,p}$, obtained via
\eq{gan-pVEV} using (\ref{PQvev12})--(\ref{PQvev13}), so that the typically unsuppressed
  ${\cal O}_{1}^{ud}$ operator dominates. In the large $m_s$ limit the complete result can be
written as
\bea
\bar{g}_{an,p} &\simeq & -\frac{G_F}{\sqrt{2}}\, \frac{8 \,c_3 \,b_0}{F_\pi^2 f_a (m_d+m_u)}\nn\\
&&{}\times \bigg\{\!\begin{array}{l}
            m_d ({\cal C}_1^{[ud]}+{\cal C}_1^{[us]})\ -m_u{\cal C}_1^{[ud]}\,b\\           
            m_d ({\cal C}_1^{[ud]}+{\cal C}_1^{[us]})\, b -m_u{\cal C}_1^{[ud]}
         \end{array},
       \label{ganp}
       \eea
where $b=(b_0+b_D+b_F)/b_0\simeq 1.2$.
A few comments on Eqs.~(\ref{gan-pVEV})\ and (\ref{ganp})\ are in order.  The chiral approach allows
us to consistently derive and account for the meson and axion tadpole contributions, thus properly
addressing interference and comparison among the various contributions.  It further includes LO
isospin-breaking effects that enter through the pion VEV (via the $b_{D,F}$ couplings) and from the
$\bar{\theta}_{\rm eff}$ term. 
Within the range of hadronic parameters here considered it
leads to a $\bar g_{ap}$ coupling about 60$\%$ larger than $\bar g_{an}$.  Finally, the results in
\eqs{gan-pVEV}{ganp} are general enough to apply to any axion model with effective RH currents,
since the model-dependent derivative axion couplings do not enter the scalar coupling.

\Sec{Experimental probes for $\bar{g}_{an,p}$.}
At present, the best sensitivity on the QCD axion exploiting axion-mediated forces is obtained by
combining limits on monopole-monopole interactions with astrophysical limits of pseudoscalar
couplings \cite{Raffelt:2012sp}.  On the other hand, monopole-dipole forces will become the best
constraining combination in laboratory experiments.  In fact, monopole-monopole interactions are
doubly suppressed in $\bar{\theta}_{\rm eff}$ while dipole-dipole forces have large backgrounds
from ordinary magnetic forces.  State of the art limits on monopole-dipole forces can be found in
Ref.~\cite{Lee:2018vaq}: the resulting lower bounds are at most at the level of
$f_a \gtrsim \sqrt{\bar{\theta}_{\rm eff}}\ 10^{13}$ GeV.

A new detection concept, by the ARIADNE collaboration \cite{Arvanitaki:2014dfa,Geraci:2017bmq},
plans to use NMR techniques to probe the axion field sourced by unpolarized Tungsten $^{184}$W
and detected by laser-polarized $^3$He.  In its current version, the experiment is sensitive to
$\bar{g}_{a^{184}\text{W}}\ g_{a^{3}\text{He}}$.  The CPV coupling axion coupling to Tungsten 
is approximated by 
$\bar{g}_{a^{184}\text{W}} \simeq 74 (\bar{g}_{ap} + \bar{g}_{ae}) + 110
\bar{g}_{an}$~\cite{Irastorza:2018dyq}, where for the QCD axion $\bar{g}_{ae} = 0$ 
at tree level. 
It is
convenient to define an average coupling to nucleons (weighting isospin breaking) as
\beq
\label{eq:bargav}
\bar{g}_{aN}\equiv \frac{74 \bar{g}_{ap} + 110 \bar{g}_{an}}{184} \, .
\eeq
The CP-conserving term, $g_{a^3\text{He}} = g_{an}$, is only sensitive to neutrons because protons
and electrons are paired in the detection sample.  Thanks to NMR, ARIADNE can
improve the sensitivity of previous searches and astrophysical limits by up to two orders of
magnitude in $(\bar{g}_{aN} g_{an})^{1/2}$ (for $m_a \in [1, 10^{4}]\,\mu$eV depending on the
spin relaxation time), before passing to a scaled-up version with a larger ${}^3$He cell reaching
liquid density.
  
To provide an example of the testing power of these future experiments, as a definite model of RH
currents we consider the paradigmatic case of the LR symmetric model (LRSM), with a PQ
symmetry.

\Sec{Application to Left Right models.}
In the minimal
LRSM~\cite{Pati:1974yy,Mohapatra:1974hk,Senjanovic:1975rk,Senjanovic:1978ev,Mohapatra:1979ia}, the
gauge group $\SU(3)_C \times \SU(2)_L \times \SU(2)_R \times \U(1)_{B-L} $ is spontaneously broken
by a scalar triplet VEV $\vev{\Delta^0_R} =v_R$ and eventually by the VEVs of a bidoublet field
$\vev{\Phi} = \diag\{v_1,e^{i\alpha}v_2\}$, where $v^2 = v_1^2 + v_2^2\ll v_R^2$ sets the
electroweak scale and $\tan\beta \equiv t_\beta = v_2/v_1$.  The single phase $\alpha$ is the source
of the new CP violation.  An important phenomenological parameter is the mixing between left and
right gauge bosons, $\zeta \simeq - e^{i \a} \sin 2 \b\,
{M_{W_L}^2}/{M_{W_R}^2}$, bound to $|\zeta| < 4\times 10^{-4}$ from direct search limits on $W_R$.

Born in order to feature the spontaneous origin of the SM parity breaking, the model is endowed with
the discrete parity $\P$, assumed exact at high scale and broken spontaneously by $v_R$. $\P$
exchanges the gauge groups, the fermion representations $Q_L$ $\leftrightarrow$ $Q_R$, and conjugates
the bidoublet $\Phi\leftrightarrow\Phi^\dag$. As a result, the Yukawa Lagrangian
$\mathcal{L}_Y= \bar Q_L (Y \Phi + \tilde Y \tilde \Phi) Q_R + \text{h.c.}$
requires hermitean $Y$, $\tilde Y$.  The diagonalization of quark masses
gives rise to a new CKM matrix $V_R$ in the $W_R$ charged currents.  Only for nonzero $\alpha$ the
masses are non-hermitean and $V_R$ departs from the standard $V_L$. An
analytical form for $V_R$ is found perturbatively in the small parameter
$y=|s_\a\, t_{2\b}|\lesssim 2m_b/m_t\simeq 0.05$~\cite{Senjanovic:2014pva,Senjanovic:2015yea}. While
the left and right mixing angles can be considered equal for our aims, $V_R$ has new external CP
phases.  For later convenience we denote them as $\theta_q$, with
$V_R={\rm diag}\{\e^{i\t_u}\!, \e^{i\t_c}\!, \e^{i\t_t}\} \, V_L\, {\rm diag}\{\e^{i\t_d}\!,
\e^{i\t_s}\!, \e^{i\t_b}\}$.  All $\theta_q$ are small deviations of $O(y)$ around $0$ or $\pi$,
corresponding to 32 physically different sign combinations of the quark mass
eigenvalues~\cite{Senjanovic:2015yea, Bertolini:2019out}.  For details on the relevant features of
the minimal LR model we refer to \cite{Senjanovic:2011zz,Bertolini:2019out} and references therein.

There are two qualitatively different ways of implementing a $\U(1)_{\rm PQ}$ symmetry in LR models,
following either the KSVZ \cite{Kim:1979if,Shifman:1979if} or the DFSZ
\cite{Zhitnitsky:1980tq,Dine:1981rt} variant. In the former, the field content of the minimal
LRSM remains uncharged under $\U(1)_{\rm PQ}$, and the pseudoscalar axion couplings to
nucleons are given by \eqs{eq:Cap}{eq:Can} with $c_q = 0$.

On the other hand, the construction of a LR-DFSZ model, with SM quarks carrying PQ charges, turns
out to be less trivial.  This is due mainly to the fact that chiral PQ charges
$\X_{Q_L} \neq \X_{Q_R}$ forbid one of the Yukawa terms in ${\cal L}_Y$,
implying unphysical mass matrices. Hence, either the LR field content must be extended
~\cite{Gu:2010zv,Gu:2010vb} (e.g.~with a second bidoublet) or effective operators must be invoked in
the Yukawa sector \cite{Dev:2018pjn,Dias:2019ezk}.  Finally, a complex singlet $\S$ to decouple the
PQ scale from $v_R$ and $v$ is needed.  A complete ultraviolet LR-DFSZ model description is not
needed here~\cite{LRDFSZinprep}, it is enough to report the axion couplings to quarks and
charged-leptons,
\beq 
\label{eq:cudDFSZ}
c_{u,\, c, \, t} = \frac{1}{3} \sin^2\beta \, , \quad 
c_{d,\, s,\, b} 
= c_{e,\, \mu,\, \tau} 
= \frac{1}{3} \cos^2\beta \, . 
\eeq

While the minimal LR model with $\P$ is a predictive theory even in the strong CP
sector~\cite{Maiezza:2014ala,Senjanovic:2020int},
the axion hypothesis can relax predictivity in the fermion as well as in the strong CP
sector, if other fields as a second bidoublet are introduced.  We stick below to the LR-KSVZ or
the LR-DFSZ case with a single bidoublet and a nonrenormalizable Yukawa term.
The axion washes
out $\bar\theta$ (and renormalizations~\cite{Maiezza:2014ala, Kuchimanchi:2014ota}), and 
observables such as e.g.\ $d_n$ and $\bar g_{an,p}$, are tightly predicted.

With this choice, quark masses set as usual a perturbativity limit on $t_\beta$,
mainly due to $m_t/m_b$: one finds $t_\beta \lesssim 0.5$~\cite{Maiezza:2010ic} or $\gtrsim 2$. The
two ranges are equivalent in the minimal model (swapping $Y$ and $\tilde Y$) but they become
physically different when the PQ symmetry acts on $\Phi$. Within this perturbative domain the
    pseudoscalar axion coupling to nucleons \eqs{eq:Cap}{eq:Can} can never vanish.

\Sec{Axion and CPV probes of LR scale.}
The RH currents in the LRSM induce the axion couplings described above.  For details on the
LRSM short-distance and the extended chiral Lagrangian we refer
to~\cite{Bertolini:2019out}.  We just recall that the short-distance
  coefficients ${\cal C}_i^{qq'}$ depend on the relevant CKM entries, carrying the additional CP
  phases of $V_R$, and on the LR gauge mixing $\zeta$. The ${\cal C}_i^{qq'}$ are renormalized at
  the 1\,GeV hadronic scale and matched with the chiral low energy constants.

To analyze the predicted $(g_{an}\bar{g}_{aN})^{1/2}$ as a function of $M_{W_R}$, we study
together the four CPV observables ($\varepsilon$, $\varepsilon'$, $d_n$, $\bar g_{aN}$), while
marginalizing on $\tan\beta$, the CP phase $\alpha$, and the 32 signs.  As in
Refs.~\cite{Bertolini:2014sua,Bertolini:2019out}, we introduce a parameter $h_i$ for each
observable, normalizing the LR contributions to the experimental central value ($\varepsilon$,
$\varepsilon'$) or upper bound ($d_n$). For the latter we take the updated 90$\%$ C.L.~result
$d_n< 1.8\times 10^{-26}\, \rm e\, cm$~\cite{Abel:2020gbr}.  The LR contributions to the indirect
CPV parameter $\varepsilon$ in kaon mixing was thoroughly analyzed in~\cite{Bertolini:2014sua} to
which we refer for details.  For the direct CPV parameter $\varepsilon'$ the latest lattice
result~\cite{Abbott:2020hxn} for the $K\to\pi\pi$ matrix element of the leading QCD penguin operator
supports the early chiral quark model prediction~\cite{Bertolini:1997ir,Bertolini:1997nf}, confirmed
by the resummation of the pion rescattering~\cite{Pallante:1999qf}, as well as more recent chiral
Lagrangian reassessments~\cite{Gisbert:2017vvj,Cirigliano:2019cpi}, including a detailed analysis of
isospin breaking.  All of the above point to a SM prediction in the ballpark of the experimental
value, albeit with a large error~\cite{Aebischer:2020jto}. We consider below two benchmark cases:
50$\%$ and 15$\%$ of $\varepsilon'$ induced by LR
physics~\cite{Bertolini:2012pu,Bertolini:2013noa}.

\begin{figure}[t]
\centering
\includegraphics[width=8cm]{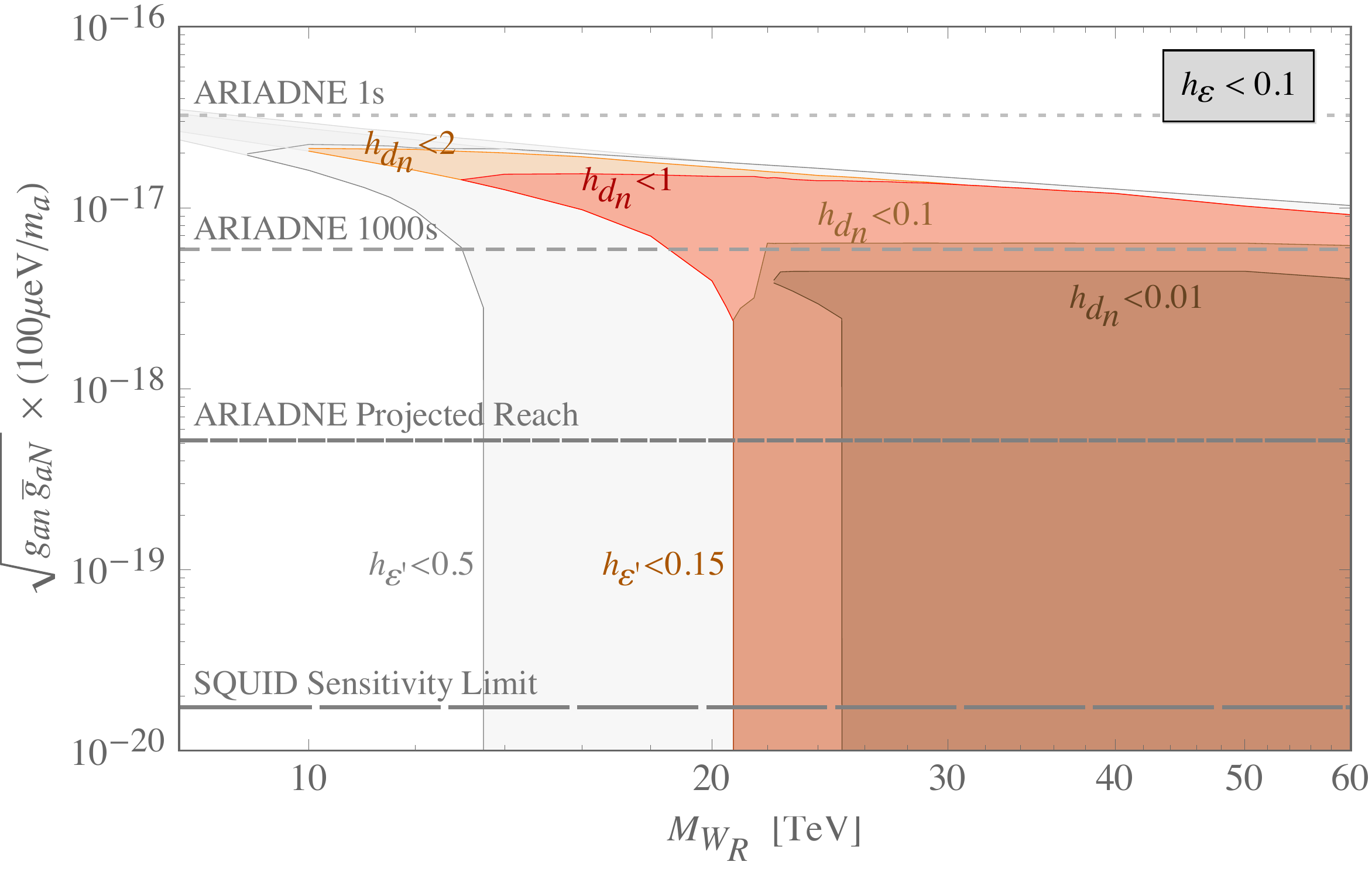}
\vspace*{-4ex}%
\caption{Regions in the LR-DFSZ model of the CPV axion nucleon coupling probed by ARIADNE.}
\label{fig:gSaNbounds_vsMWR}       
\end{figure}

The average CPV nucleon coupling in \eq{eq:bargav} is computed using \eq{ganp}.  With the updated $d_n$
bound and including the strange quark contributions, 
we obtain
\bea
  \bar g_{aN} &=& \frac{|\zeta|}{10^{-5}}
   \Big[ 6.4 \sin \alpha_{ud}
   + 0.7 \sin\alpha_{us}\Big] \frac{m_a}{100 \, \text{$\mu $eV}} 10^{-12} \nn\\[.4ex] 
  h_{d_n}&=&  \frac{|\zeta|}{10^{-5}}
   \Big[7.1\sin \alpha_{ud}
   -3.4   \sin \alpha_{us}\Big]
   \label{eq:numdn}\nn\\[.6ex] 
   h_{\varepsilon'}&=& \frac{|\zeta|}{10^{-5}} \Big[9.2 \sin
   \alpha_{ud} + 9.2 \sin \alpha_{us}\Big]\,,
\label{eq:numepsprime}
\eea
where $\alpha_{qq'}=\alpha-\t_q-\t_{q'}$. We recall that all phases $\t_q$ depend on a single
parameter. Also, $\alpha_{ud}\simeq\alpha_{us}$ modulo $\pi$ for $M_{W_R}\lesssim 30\,$TeV from the
$h_\eps$ constraint~\cite{Bertolini:2014sua}, which plays an important role in enforcing a
  tight correlation between the above observables.  The subleading role of the Cabibbo
  suppressed $us$ Wilson coefficient in $\bar g_{aN}$ is clear, unlike the case of $d_n$ where
  the leading $ud$ contribution is canceled as mentioned above~\cite{Bertolini:2019out}.

The model-dependent pseudoscalar coupling $g_{an}$ in the monopole-dipole interaction is taken for
the LR-DFSZ case via \eq{eq:cudDFSZ}.  Similar results are obtained for LR-KSVZ, for which however
$g_{an}$ is compatible with zero,~\eq{eq:Can}.

In \fig{fig:gSaNbounds_vsMWR} we show the allowed regions of $(g_{an}\bar{g}_{aN})^{1/2}$ as a
function of $M_{W_R}$, together with the reach of three phases of ARIADNE ($1$s, $1000$s,
projected)~\cite{Arvanitaki:2014dfa,Geraci:2017bmq} and the SQUID sensitivity limit.  We scale the
coupling combination by $f_a\propto 1/m_a$, making the prediction independent from it. With this
normalization the experiment sensitivities vary mildly with $m_a$, and we show their best reach,
attained for $m_a\sim10^{2\div3} \,\mu$eV.  Present limits from astrophysics \cite{Raffelt:2012sp}
and monopole-dipole experiments \cite{Lee:2018vaq} lie above the plot and are hence ineffective to
probe the LR scale.

The predicted regions depend on the constraints on $h_{\eps}$, $h_{\eps'}$ and $h_{d_n}$. In the
colored area the LR contribution to $\eps'$ is allowed up to 15\%, while in light gray we relax it
to 50\%, given the present theoretical uncertainties.  In either case, a lower bound on $\bar g_{aN}$
arises, for $M_{W_R}\lesssim 20$ or $13\,$TeV respectively.
The origin of this lower bound is traced to the fact that, in the LRSM with $\P$, for a few TeV
$M_{W_R}$ the CPV effects cannot be eliminated by taking $\a\to 0$: an exceedingly large
contribution to $h_\eps$ would remain from the CKM phase in $V_R$, thus a destructive interference
from additional CP phases is required~\cite{Bertolini:2014sua}.
Thus, for instance, a positive detection from ARIADNE below $2\times 10^{-18}$ with
$m_a\approx 100 \,\mu$eV would falsify such TeV-scale LR-DFSZ scenario.  Instead, a measurement
above $10^{-17}$ would result in a rejection of the LR-DFSZ model or a sharp upper bound on
$M_{W_R}$, at the reach of a future collider.

Given the square root in $(g_{an}\bar{g}_{aN})^{1/2}$, the probed observable depends mildly on the
new physics scale. Indeed, the upper boundary of the shaded region decreases as $ 1/M_{W_R}$, and we
find that within the ARIADNE sensitivity the model provides possible signals up to
$M_{W_R}\sim 1000\,$TeV.  Standard flavour observables, decoupling as $1/M_{W_R}^2$, have a more
limited reach.

The effect of the present and future constraints on $d_n$ are shown with increasingly darker
shadings, from a most conservative $h_{d_n}<2$ (accounting for hadronic uncertainties), to a most
stringent future bound of $h_{d_n}<0.01$.  The bounds on $d_n$ limit from above the predicted
axion-mediated force.  For instance $h_{d_n}<0.1$ implies a prediction at the level of the ARIADNE
1000s sensitivity.

To conclude, we provided a complete and consistent calculation of the CPV axion couplings to
matter and applied it to the case RH currents, showing that axion-mediated forces provide a
powerful probe of the CPV structure and scale of minimal LR-PQ scenarios. It is amusing that the
first hints of high-energy parity restoration may possibly be revealed in a condensed matter lab.

\begin{acknowledgments}
\Sec{Acknowledgments.} The work of LDL is supported by the Marie Sk\l{}odowska-Curie 
Individual Fellowship grant AXIONRUSH (GA 840791) 
and the Deutsche Forschungsgemeinschaft under Germany's Excellence Strategy 
- EXC 2121 Quantum Universe - 390833306. 
The work of F.N.~was partially supported by the research grant No.~2017X7X85K 
under the program PRIN 2017 funded by the Ministero dell'Istruzione, 
Universit\`a e della Ricerca (MIUR)
\end{acknowledgments}

\bibliographystyle{utphysmod.bst}
\bibliography{axionLR}

\end{document}